\documentclass[twocolumn,showpacs,preprintnumbers,amsmath,amssymb,prl]{revtex4}

\usepackage{graphicx}
\usepackage{bm}

\begin{document}

\preprint{}

\title{Absence of kinetic effects in reaction-diffusion processes in scale-free networks}
\author{Lazaros K. Gallos}
\author{Panos Argyrakis}
\affiliation{Department of Physics, University of Thessaloniki, 54124 Thessaloniki, Greece}

\date{\today}

\begin{abstract}
We show that the chemical reactions of the model systems of A+A$\rightarrow$0 and A+B$\rightarrow$0
when performed on scale-free networks exhibit drastically different behavior as compared to the same
reactions in normal spaces. The exponents characterizing the density evolution as a function of time
are considerably higher than 1, implying that both reactions occur at a much faster rate.
This is due to the fact that the discerning effects of the generation
of a depletion zone (A+A) and the segregation of the reactants (A+B) do not occur at all as in
normal spaces. Instead we observe the formation of clusters of A (A+A reaction) and of mixed
A and B (A+B reaction) around the hubs of the network. Only at the limit of very sparse networks
is the usual behavior recovered.
\end{abstract}

\pacs{82.20.-w, 05.40.-a, 89.75.Da, 89.75.Hc}

\maketitle

The model bimolecular chemical reactions of the A+A and A+B type have been heavily studied since the
pioneer work of Ovchinnikov and Zeldovich \cite{ZO78}, postulating a behavior drastically different
than the mean field predictions, especially in low dimensions and in fractal structures.
These were verified additionaly by numerical simulations, which showed pictorially the dominating
effect in these two systems, i.e. the generation of the depletion zone for the A+A type shown by
Torney and McConnell \cite{TM83}, and the spatial segregation
of the two types of reactants in the A+B type shown by Toussaint and Wilczek \cite{TW83},
In the past twenty years a large number of works
extended these initial ideas and explained in detail how these effects come about, showed the
existence of several temporal regimes, explained the crossovers between the early time and
long time behavior, and a wealth of other information, rendering these systems as some of the
most heavily studied systems of interacting particles \cite{benBook}.

Classically, the mean-field prediction for the density $\rho$ of the surviving particles in
A+A and equimolar A+B type reactions is
\begin{equation}
\frac{1}{\rho}-\frac{1}{\rho_0} = kt^f \,.
\end{equation}
where $f=1$, $k$ is the rate constant, and $\rho_0$ is the particle density at $t=0$.
In non-classical kinetics, though, an `anomalous' behavior has been observed. In a space
with dimensionality $d$ the exponent $f=d/d_c$ for $d\leq d_c$, and $f=1$ for $d>d_c$. The upper critical dimension $d_c$ equals
2 for A+A, and $d_c=4$ for A+B.
Similarly, anomalous behavior has been observed when the diffusion-reaction process takes place on
different geometries, such as on fractals \cite{bAH87} where $f=d_s/d_c$, or on dendrimer structures \cite{AK00}.
In all these cases the limiting value of the exponent is $f=1$, valid even in infinite dimensions.

In this Letter we study reactions of the A+A$\rightarrow$0 and A+B$\rightarrow$0 type taking place on scale-free networks.
Networks of this type have been recently found to characterize a wide range of systems in
nature and society \cite{AB02, DM02}, including the Internet, the WWW, chemicals linked via chemical reactions,
sexual contacts, ecological systems, etc. The term scale-free refers to the absence of a characteristic scale
in the connectivity of the nodes comprising the network. Thus, each node of the system
has $k$ links to other nodes in the system with a probability
\begin{equation}
P(k) \sim k^{-\gamma} \, ,
\end{equation}
where $\gamma$ is a parameter characteristic of the structure of the network.
Although these systems are very large
their diameter is usually small, a property which is usually referred to as the `small-world effect'.
The topology of such a network is quite complex and leads to a drastically different behavior for
the above mentioned chemical reactions.

Due to the different structure, as compared to lattices, it is of interest to examine a reaction-diffusion
process on a scale-free network and observe the evolution of the two major effects, i.e. generation
of the depletion zone and segregation of the reactants. Actual situations could involve a virus-antivirus reaction
of the Internet, or epidemics in a social network. Thus, we performed Monte-Carlo calculations
of these model systems on a scale-free network and monitored the usual parameters, i.e.
density as a function of time, and the exponent $f$ (equation 1).

A scale-free network is constructed as follows. For a given
$\gamma$ value we fix the number
of nodes $N=10^6$ and we assign the degree $k$ (number of links) for each node by drawing a
random number from a power law distribution $P(k)\sim k^{-\gamma}$.
We do not impose any cutoff value for the maximum nodes connectivity, so that $k$ can assume
any value in the range [1,N].
We then randomly select and connect pairs of links between nodes that have not yet 
reached their preassigned connectivity
and have not already been directly linked to each other.
Finally, we isolate the largest cluster formed in the network. This is identified via a burning
algorithm, i.e. we start from a randomly chosen node, assign an index i to it, and `burn' all
of its neighbors, by assigning the same index i to them. The process repeats iteratively and
when there are no more neighbors to be burnt in this cluster, a new non-burnt node is chosen again,
an index i+1 is assigned to it, etc, until all nodes have been visited with this process. The largest
cluster can now be easily identified by counting the number of nodes with the same index.

\begin{figure}
\includegraphics{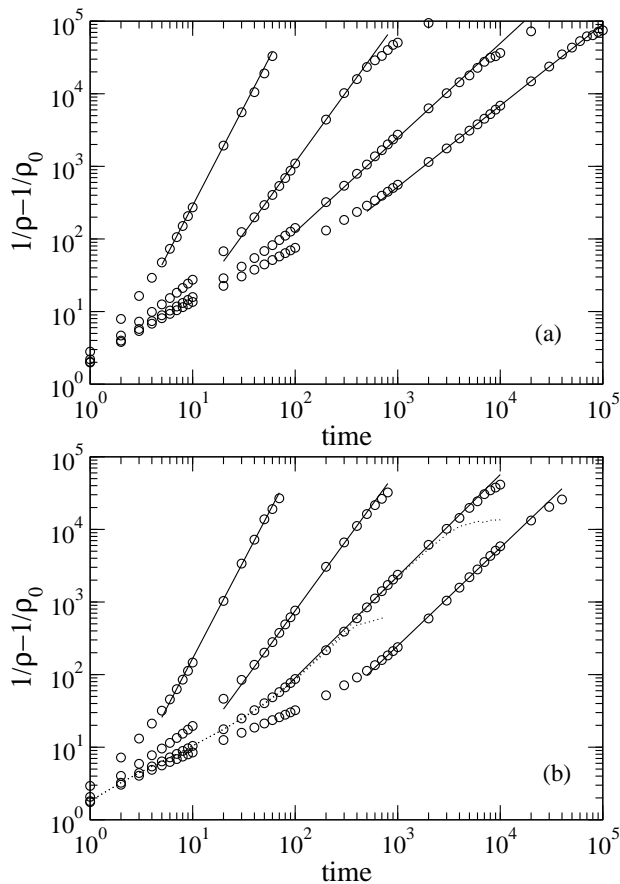}
\caption{\label{fig1} Plots of the reaction progress $1/\rho-1/\rho_0$, as a function of
time for (a) the A+A$\rightarrow$0 reaction and (b) the A+B$\rightarrow$0 reaction on scale-free
networks of (left to right) $\gamma=$2.0, 2.5, 3.0, and 3.5. The initial density was $\rho_0=0.5$
and $\rho_0=0.25$ respectively. All results correspond to networks of $N=10^6$ nodes, except for
the dotted lines in (b), where we present results for systems of $N=10^4$ and $10^5$ nodes.
The symbols represent the simulation data, while continuous lines are the asymptotic best-fit lines.}
\end{figure}

The reaction mechanism is similar to the one routinely used in the literature for the case of
regular lattices. For the A+A$\rightarrow$0 reaction, an initial fraction of the nodes $\rho_0$
of the largest cluster
is randomly chosen. These nodes are occupied by A particles. One particle is chosen at random
and the direction of its move is also chosen at random with equal probability to one of its
neighbor sites (nodes directly linked to its
present position). The total time advances by $1/N$. If this node is already occupied by
another particle the two A particles are annihilated, otherwise the particle is moved to the
new node. For the A+B$\rightarrow$0 reaction, an equal initial fraction $\rho_0$ of A and B particles
are randomly placed on the network. Particles move using the same algorithm described above, but now
when a particle tries to move to a node where a particle of the same type resides the move is not
allowed, although the time advances. When an A particle encounters a B particle these two particles
annihilate. We monitor the population of particles as a function of time.

\begin{figure}
\includegraphics{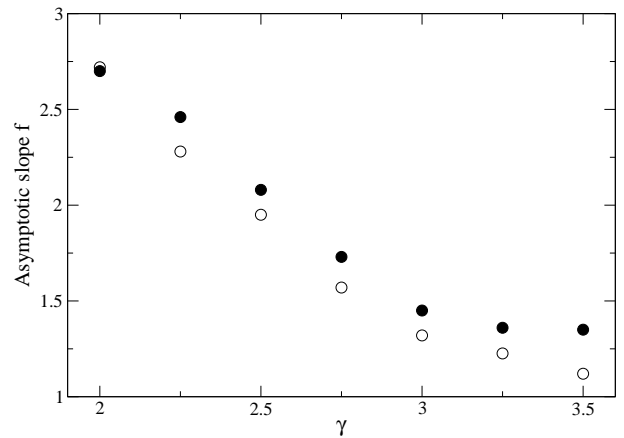}
\caption{\label{fig2} Asymptotic reaction rates (slopes of the lines in figure 1), as a function of $\gamma$,
for the A+A$\rightarrow$0 reaction (open symbols) and the A+B$\rightarrow$0 reaction (filled symbols).}
\end{figure}

In figure 1 we present the evolution of the particle concentration as a function of time for the
(a) A+A and (b) A+B reactions. Results for networks with different $\gamma$ values are presented,
where $\gamma$ ranges from 2 to 3.5, representing a varying degree of nodes connectivities (from dense to
sparse networks). As we can
see, the curves follow power-law behavior with two distinct regimes. In all cases, there exists a crossover
between the early-time regime and the asymptotic limit. The location of the crossover point increases with
$\gamma$. For $\gamma=2$ the crossover takes place as early as 10 steps or less, while for $\gamma=3.5$
it is of the order of 1000 steps. The reaction rate, also, is much faster for lower $\gamma$ values,
with the concentration falling to $10^{-4}$ in only 50 steps for $\gamma=2$, for both types of reaction.
In figure 1b and for $\gamma=3$ we also give results for smaller networks ($N=10^4$ and $N=10^5$),
in order to ascertain that there are no finite-size effects for the calculation of the asymptotic slopes.

\begin{figure}
\includegraphics{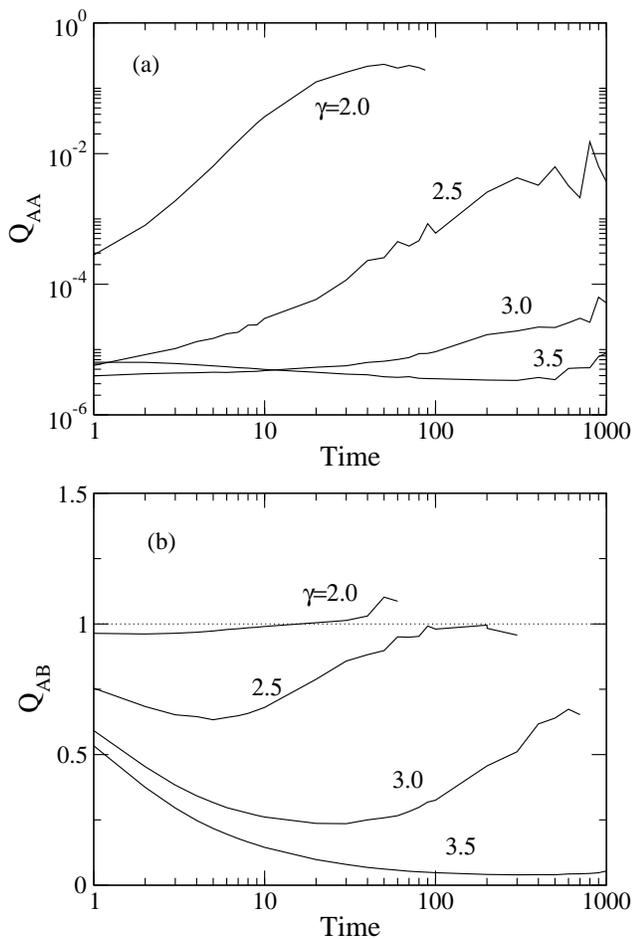}
\caption{\label{fig3} (a) A+A reaction: Percentage $Q_{AA}$ of contacts between A particles over the total number of possible contacts
$N_{AA}/ N(N-1)$ as a function of time.
(b) A+B reaction: Percentage of AB contacts over (AA+BB) contacts
as a function of time. The line at $Q_{AB}=1$ corresponds to complete mixing. The $\gamma$ values are as marked.}
\end{figure}

These observations are in contrast with reaction schemes on regular lattices. This can be clearly
seen in the calculation of the asymptotic slopes for the curves of figure 1. These slopes
(corresponding to the exponent $f$ of equation 1) are presented in figure 2, as a function of $\gamma$.
There is a small difference between the slopes of the A+A and the A+B types (with the latter slightly
larger), but in all cases the slope is greater than 1. The exponent $f$ can acquire very large values
($f=2.75$ for $\gamma=2$) and is a monotonically decreasing function of $\gamma$. As already stated,
on all other geometries studied in the literature, the exponent $f$ is lower than 1, while on scale-free
networks we see that the value $f=1$ is reached only asymptotically for networks with large $\gamma$.

In order to understand the spatial distribution of the particles, we calculate
the number of close contacts they form (we consider a close contact as the existence of a link between two network nodes).
In the case of the A+A reaction, for fixed time, we measure the fraction $Q_{AA}$
defined as the number of contacts $N_{AA}$ between A particles over the total possible number
of contacts, i.e.
\begin{equation}
Q_{AA} = \frac{N_{AA}}{N(N-1)} \,.
\end{equation}
The value of $Q_{AA}=1$ corresponds to the extreme case of all particles forming one cluster,
while a decrease of this value indicates the existence of a depletion zone (particles are placed
apart from each other).
In figure 3a we calculate the fraction $Q_{AA}$ for different networks as a function of time.
When $\gamma=2$ or 2.5 a depletion zone initially forms, but immediately the number of contacts
increases with time. Most of the particles
are clustered in a small region and as time advances this clustering increases, i.e. particles continue
to gather in positions close to each other, in the vicinity
of the most connected nodes (hubs). This is in striking contrast to the well-established formation of
a depletion zone on regular lattices. For $\gamma=3$ we can see that the number of contacts
remains almost constant with time, while for $\gamma=3.5$ we can recover the formation of the
usual depletion zone, where the number of contacts decreases monotonically with time.

Similar conclusions are found for the A+B reaction (figure 3b).
In this case we measure the number of contacts between unlike particles
compared to the number of contacts between particles of the same type.
We use the ratio $Q_{AB}$ defined as \cite{NK88}:
\begin{equation}
Q_{AB} = \frac{N_{AB}}{N_{AA}+N_{BB}} \,.
\end{equation}
The value of $Q_{AB}$ tends to 0 when segregation of like species is formed, while it tends to a value of 1 when
complete mixing of A and B occurs.
When $\gamma=2$ there is clearly no segregation of like particles, even at early times,
but both A and B particles mix together. For $\gamma=2.5$ and $\gamma=3.0$ there are hints of segregation,
since the number of unlike particles contacts decreases compared to 
like particle contacts, but soon particles start to aggregate. The time needed for the transition to mixing and the
extend of the depletion zone increase with $\gamma$. At $\gamma=3.5$ the classical picture with a clear
segregation of particles reappears,
where like particles gather in clusters with a scarce presence of unlike particles at all times.

\begin{figure}
\includegraphics{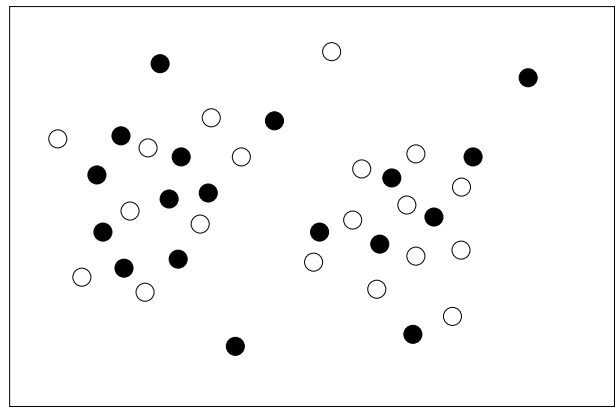}
\caption{\label{fig4} Lattice analogue pictorial of the particles configuration in space for the A+B$\rightarrow$0
reaction. Notice that particles of different kind gather in the same region, in the close vicinity of the hubs.}
\end{figure}

This peculiar behavior can be attributed to the structural characteristics of the scale-free networks.
Such networks are known to heavily rely on the existence of hubs, which are nodes with a large number of connections.
Moreover, such networks have been shown to have an extremely small diameter, of
the order $\ln(\ln N)$ \cite{CH03}. These two factors can explain the results of the present work.
The small diameter of the network causes the majority of the particles to be at a close distance to
each other, and most of the nodes can reach a hub through a small number of links. Thus, we expect
that diffusion will very soon bring the particles close to each other and they will react within a
short period of time. This is indeed the case of figure 1, where we can see that the process ends rapidly.
The importance of the hubs diminishes as we increase $\gamma$, since now nodes are less connected to each
other, and the familiar notions of depletion zones and
segregation are restored for $\gamma=3.5$. One can picture this process as a biased
walk of the particles towards the hubs, where the probability of encounter of another particle is much higher
than in the case of a regular lattice walk.

These concepts correspond to the idea of continuous mixing. For the A+A reaction and low $\gamma$ values
there exists no depletion zone. This is due to the fact that neighboring A particles annihilate initially
in a very fast rate, but as time advances the diffusion of the A particles brings them close to the hubs,
which dominate the structure. Thus, there are always pairs of A close to each other (absence of depletion).
For larger $\gamma$ values, i.e. when the networks are more sparse, the connectivity of the hubs
diminishes, and the regular result of a depletion zone is reproduced.

In figure 4 we present a lattice analogue pictorial
of the A+B case. Although unlike particles that were close at $t=0$ have been annihilated,
at longer times there is still no segregation, since all particles, independently of their type, move
towards the hubs. We thus have the formation of clusters made of A and B particles, which are located
close to the hubs. This means that A and B particles are always mixed because of the underlying structure
and the reaction advances at a constant rapid rate. We have indeed verified in our simulations that
the majority of the annihilation reactions take place on the hubs. Notice also that diffusion on scale-free
networks is non-Markovian, since the walk depends on the exact particle location, due to the largely
varying connectivity of the nodes.

A similar situation is encountered in another system, that of L\'{e}vy walks, where it was observed
that the rare long-distance jumps break the formation of the segregated regions, acting in effect like
a stirring mechanism \cite{ZK94,ZKS96}.

Summarizing, we have presented evidence that reaction-diffusion processes in scale-free networks are
different in their nature compared to lattice models, exhibiting rapid reaction rates ($f>1$).
This is due to the small diameter of the networks and the existence of the hubs. These
differences are more pronounced in compact networks of low $\gamma$ values, while for sparse
networks, e.g. $\gamma\geq 3.5$, the behavior is the same as for regular lattices.


\begin{thebibliography}{}

\bibitem{ZO78} A.A. Ovchinnikov and Ya.B. Zeldovich, Chem. Phys. {\bf 28}, 215 (1978).
\bibitem{TM83} D.C. Torney and H.E. McConnell, J. Phys. Chem. {\bf 87}, 1441 (1983);
Proc. R. Soc. Lond. A {\bf 387}, 147 (1983).
\bibitem{TW83} D. Toussaint and F. Wilczek, J. Chem. Phys. {\bf 78}, 2642 (1983).
\bibitem{benBook} D. ben-Avraham and S. Havlin, \emph{Diffusion and reactions in fractals and disordered systems}
(Cambridge University Press, Cambridge, 2000).
\bibitem{bAH87} D. ben-Avraham and S. Havlin, Adv. in Phys. {\bf 36}, 695 (1987).
\bibitem{AK00} P. Argyrakis and R. Kopelman, Chem. Phys. {\bf 261}, 391 (2000).
\bibitem{AB02} R. Albert and A.-L. Barabasi, Rev. Mod. Phys. {\bf 74}, 47 (2002).
\bibitem{DM02} S.N. Dorogovtsev and J.F.F. Mendes, Adv. Phys. {\bf 51}, 1079 (2002).
\bibitem{NK88} J.S. Newhouse and R. Kopelman, J. Phys. Chem. {\bf 92}, 1538 (1988).
\bibitem{CH03} R. Cohen and S. Havlin, Phys. Rev. Lett. {\bf 90}, 058701 (2003).
\bibitem{ZK94} G. Zumofen and J. Klafter, Phys. Rev. E {\bf 50}, 5119 (1996).
\bibitem{ZKS96} G. Zumofen, J. Klafter and M.F. Shlesinger, Phys. Rev. Lett. {\bf 77}, 2830 (1996).

\end{thebibliography}
\end{document}